\documentclass{ecai2008}
\pdfoutput=1
\usepackage{graphicx}
\usepackage{xspace}
\usepackage{amssymb}
\usepackage{url}
\usepackage{times}
\usepackage{latexsym}

\newcommand{\eg}{e.\,g.,\xspace}
\newcommand{\ie}{i.~e.,\xspace}
\newcommand{\del}{del.icio.us\xspace}

\newenvironment{definition}[1][Definition]{\begin{trivlist}
\item[\hskip \labelsep {\bfseries #1}]}{\end{trivlist}}

\newcommand{\R}{\mathbb{R}}

\newcommand{\card}{\mathrm{card}}

\begin{document}

\title{Semantic Analysis of Tag Similarity Measures in Collaborative Tagging Systems}

\author{Ciro Cattuto\institute{Complex Networks Lagrange Laboratory (CNLL), ISI 
Foundation, 10133 Torino, Italy, email:cattuto@isi.it} \and Dominik Benz$^2$ \and 
Andreas Hotho$^2$ \and Gerd Stumme\institute{Knowledge \& Data Engineering Group, 
University of Kassel, 34121 Kassel, Germany, email: \{benz,hotho,stumme\}@cs.uni-kassel.de} }

\maketitle

\begin{abstract}
Social bookmarking systems allow users to organise collections of resources on the 
Web in a collaborative fashion. The increasing popularity of these systems as well as
first insights into their emergent semantics have made them relevant to disciplines 
like knowledge extraction and ontology learning. The problem of devising
methods to measure the semantic relatedness between tags 
and characterizing it semantically is still largely open.
Here we analyze three measures of tag relatedness: tag co-occurrence,
cosine similarity of co-occurrence distributions, and FolkRank,
an adaptation of the PageRank algorithm to folksonomies.
Each measure is computed on tags from a large-scale dataset crawled
from the social bookmarking system del.icio.us. To provide a semantic 
grounding of our findings, a connection to WordNet (a semantic lexicon for the
English language) is established by mapping tags into synonym sets of WordNet,
and applying there well-known metrics of semantic similarity. Our results clearly expose
different characteristics of the selected measures of relatedness,
making them applicable to different subtasks of knowledge extraction
such as synonym detection or discovery of concept hierarchies.
\end{abstract}

\vspace{-0.1cm}

\section{Introduction}
\label{sec-introduction}

Social bookmarking systems have become extremely popular in recent years. 
Their underlying data structures, known as \emph{folksonomies},
consist of a set of users, a set of free-form keywords
(called \emph{tags}), a set of resources, and a set of tag assignments,
\ie a set of  user/tag/resource triples. As folksonomies are large-scale
bodies of lightweight annotations provided by humans, they are becoming
more and more interesting for research communities that focus on extracting 
machine-processable semantic structures from them.
The structure of folksonomies, however, differs fundamentally from that
of e.g., natural text or web resources, and sets new challenges
for the fields of knowledge discovery and ontology learning.
Crucial hereby are the concepts of similarity  and relatedness.
Here we will focus on similarity and relatedness of tags,
because this affords comparison with well-established measures
of similarity in existing lexical databases.

Ref.~\cite{budanitsky06evaluating} points out that similarity
can be considered as a special case of relatedness.
As both similarity and relatedness are semantic notions,
one way of defining them for a folksonomy is to map the tags
to a thesaurus or lexicon like Roget's thesaurus\footnote{\url{http://www.gutenberg.org/etext/22}}
or WordNet~\cite{fellbaum98wordnet}, and to measure the relatedness there
by means of well-known metrics. The other option is to define
measures of relatedness directly on the network structure of the folksonomy.
There are several obvious possibilities and most of them use
statistical information about different types of co-occurrence
between tags, resources and users. Another possibility is to adopt the
\emph{distributional hypothesis}~\cite{firth57synopsis,harris68mathematical}, 
which states that words found in similar contexts
tend to be semantically similar. One important reason for using
distributional measures in folksonomies instead of mapping tags
to a thesaurus is the observation that the vocabulary of folksonomies
includes many community-specific terms which did not make it yet into
any lexical resource.

The distributional hypothesis is also at the basis of a number of approaches
to synonym acquisition from text corpora \cite{ciminao06phd}.
As in other ontology learning scenarios, clustering techniques are often 
applied to group similar terms extracted from a corpus, and a core
building block of such procedure is the metric used to judge term similarity.
In order to adapt these approaches to folksonomies, several distributional
measures of tag relatedness have been used in theory or implemented in applications~\cite{heymann06collaborative, schmitz2006inducing}. 
In most studies, however, the selected measures of relatedness
seem to have been chosen in a rather ad-hoc fashion.
We believe that a deeper insight into the semantic properties
of relatedness measures is an important prerequisite
for the design of ontology learning procedures that are capable
of successfully harvesting the emergent semantics of a folksonomy.

In this paper, we consider the three following measures for the 
relatedness of tags: the \emph{co-occurrence count}, the \emph{cosine 
similarity}~\cite{salton89} of co-occurrence distributions,
and \emph{FolkRank}~\cite{hotho06information}, a graph-based measure 
that is an adaptation of PageRank~\cite{page98pagerank} to folksonomies.
Our analysis is based on data from a large-scale snapshot
of the popular social bookmarking system \del\footnote{\url{http://del.icio.us/}}.
To  provide a semantic grounding of our folksonomy-based measures,
we map the tags of \del to synsets of WordNet
and use the semantic relations of WordNet to infer corresponding semantic
relations in the folksonomy. In WordNet, we measure the similarity
by using both the taxonomic path length and a similarity measure
by Jiang and Conrath~\cite{jiang97semantic} that has been validated
through user studies and applications~\cite{budanitsky06evaluating}.
The use of taxonomic path lengths, in particular, allows us to inspect
the edge composition of paths leading from one tag to the corresponding
related tags, and such a characterization proves to be especially insightful.

The paper is organized as follows: In the next section, we discuss
related work. In Section~\ref{sec-data} we provide a definition
of folksonomy and describe the del.icio.us data on which our experiments are based.
Section~\ref{sec-measures-I} describes the three measures of relatedness
that we will analyze. Section~\ref{sec-measures-II} provides first
examples and qualitative insights.
The semantic grounding of the
measures in WordNet is described in Section~\ref{sec-grounding}.
We discuss our results in the context of ontology learning in
Section~\ref{sec-discussion}, where we also point to future work.

\section{Related Work}
 \label{sec-relatedwork}

One of the first scientific publications about folksonomies is
\cite{mathes04folksonomies}, where several concept of bottom-up social
annotation are introduced. Ref.~\cite{lambiotte05tripartite,
mika2005ontologies} introduce a tri-partite graph representation
for folksonomies, where nodes are users, tags and resources.
Ref.~\cite{golder2006structure} provides a first quantitative analysis
of \del.

A considerable number of investigations is motivated by the vision of
``bridging the gap'' between the Semantic Web and Web 2.0
by means of ontology-learning procedures based on folksonomy annotations. 
Ref.~\cite{mika2005ontologies} provides a model of semantic-social networks
for extracting lightweight ontologies from \del. Other approaches for
learning taxonomic relations from tags 
are \cite{heymann06collaborative,schmitz2006inducing}. 
Ref.~\cite{Halpin_et_al_2006} presents a generative model for folksonomies
and also addresses the learning of taxonomic relations.
Ref.~\cite{zhang06-emergent} applies 
statistical methods to infer global semantics from a folksonomy.
The distribution of tag co-occurrence frequencies has been 
investigated in~\cite{cattuto2007pnas} and the network structure 
of folksonomies was investigated in~\cite{cattuto2007aicomm}. 

After comparing distributional measures on natural text with measures 
for semantic relatedness in thesauri like WordNet, 
\cite{mohammadSubmittedDistributional} concluded that
``distributional measures [\dots] can easily provide domain-specific similarity measures for a large number of domains [\dots].''
Our work presented in this paper indicates that 
these findings can be transferred to folksonomies.

\section{Folksonomy Definition and Data}
 \label{sec-data}

In the followin we will use the definition of
folksonomy provided in~\cite{hotho06information}:
\begin{definition}
 \label{def-folksonomy}
A \emph{folksonomy} is a tuple $\mathbb{F}:=(U,T,R,Y)$ where $U$,
$T$, and $R$ are finite sets, whose elements are called
\emph{users}, \emph{tags} and \emph{resources}, respectively., and $Y$ is a
ternary relation between them, \ie $Y\subseteq U\times
  T \times R$. A \emph{post} is a triple $(u, T_{ur}, r)$ with $u\in U$, $r\in $,
and $T_{ur}:=\{t\in T\mid (u,t,r)\in Y\}$.
\end{definition}

Users are typically represented by their user ID, tags may be arbitrary 
strings, and resources depend on the system and are usually represented
by a unique ID.

For our experiments we used data from the social bookmarking system
del.icio.us, collected in November 2006.
As one main focus of this work is to characterize tags
by their distribution of co-occurrence with other tags,
we restricted our data to the 10,000 most frequent tags of del.icio.us,
and to the resources/users that have been associated with at least
one of those tags.
One could argue that tags with low frequency have a higher
information content in principle --- but their inherent sparseness of 
co-occurrence makes them less useful for the study of distributional measures.
The restricted folksonomy consists of $|U|=476,378$ users, $|T|=10,000$ 
tags, $|R|=12,660,470$ resources, and $|Y|=101,491,722$ tag assignments.

\section{Measures of Relatedness}
 \label{sec-measures-I}

A folksonomy can be also regarded as an
undirected tri-partite hyper-graph $G = (V, E)$, where $V = U
\cup T \cup R$ is the set of nodes, and $E = \{
\{u,t,r\} \mid (u,t,r) \in Y\}$ is the set of hyper-edges.
Alternatively, the folksonomy hyper-graph can be represented
as a three-dimensional (binary) adjacency matrix. In Formal Concept
Analysis~\cite{Ganter99} this structure is known as a \emph{triadic
context}~\cite{LehmannWille95}. All these equivalent notions make
explicit that folksonomies are special cases of three-mode data.
Since measures for similarity and relatedness are not well developed
for three-mode data yet, we will consider two- and one-mode views on the data.
These two views will be complemented by
a graph-based approach for discovering related tags (FolkRank)
which makes direct use of the three-mode structure.

\subsection*{Co-Occurrence}

Given a folksonomy $(U,T,R,Y)$, we define the \emph{tag-tag
co-occurrence graph} as a weighted, undirected graph, whose set of
vertices is the set $T$ of tags, and where two tags $t_1$ and $t_2$
are connected by an edge, iff there is at least one post
$(u,T_{ur},r)$ with $t_1,t_2\in T_{ur}$. The \emph{weight} of this
edge is given by the number of posts that contain both $t_1$ and
$t_2$,
\ie
 \begin{equation}
 w(t_1,t_2):=\card\{(u,r)\in U\times R
\mid t_1,t_2\in T_{ur}\}\enspace.
 \end{equation}

Co-occurrence relatedness between tags is given directly by the edge weights.
For a given tag $t \in T$, the tags that are most related to it are
thus all tags $t'\in T$ with $t'\neq t$ such that $w(t,t')$ is
maximal. In the sequel, we will denote the co-occurrence relatedness
also by \emph{freq}.

\subsection*{Cosine Similarity}

We introduce a distributional measure of tag relatedness by
computing the cosine similarity of tag-tag co-occurrence distributions.
Specifically, we compute the cosine similarity~\cite{salton89}
in the vector space $\R^T$, where each tag $t$ is represented by a vector
$\vec{v}_t\in\R^T$ with $v_{tt'}:=w(t,t')$ for $t\neq t'\in T$ and
$v_{tt}=0$. The reason for giving weight zero between a node
and itself is that we want two tags to be considered related when
they occur in a similar context, and not when they occur together.

If two tags $t_1$ and $t_2$ are represented by
$\vec{v}_1,\vec{v}_2\in \R^n$, then their cosine similarity is
defined as:
\begin{equation}
\mathrm{cossim}(t_1,t_2):=\arccos\measuredangle(\vec{v}_1,\vec{v}_2)=
\frac{\vec{v}_1\cdot\vec{v}_2}{||\vec{v}_1||_2\cdot||\vec{v}_2||_2}\enspace
\end{equation}

\subsection*{FolkRank}
 \label{subsub-folkrank}

The PageRank algorithm~\cite{brin98anatomy} reflects the idea that a web page 
is important if there are many pages linking to it, and if those pages are 
important themselves. The same principle 
was employed for folksonomies in~\cite{hotho06information}: a resource which is 
tagged with important tags by important users becomes important itself. The 
same holds, symmetrically, for tags and users. By modifying the weights for a 
given tag in the random surfer vector, FolkRank can compute a ranked list of 
relevant tags. Ref.~\cite{hotho06information} provides a detailed description.

\section{Qualitative insights}
\label{sec-measures-II}

\begin{table*}
\centering
 \caption{Examples of most related tags measured by co-occurrence}
 \label{table-examples-cooc}
\begin{tabular}{|c|c||c|c|c|c|c|} \hline rank & tag & 1 &
2 & 3 & 4 & 5 \\ \hline\hline 13 & web2.0 & ajax & web & tools &
blog & webdesign \\ \hline 15 & howto & tutorial & reference & tips
& linux & programming \\ \hline 28 & games & fun & flash & game &
free & software \\ \hline 30 & java & programming & development &
opensource & software & web \\ \hline 39 & opensource & software &
linux & programming & tools & free \\ \hline 1152 & tobuy & shopping
& books & book & design & toread \\ \hline
\end{tabular}
\vspace{1cm}
\centering
 \caption{Examples of most related tags measured by cosine
similarity}
 \label{table-examples-cosine}
\begin{tabular}{|c|c||c|c|c|c|c|} \hline rank & tag & 1 &
2 & 3 & 4 & 5 \\ \hline\hline 13 & web2.0 & web2 & web-2.0 & webapp
& ``web & web\_2.0 \\ \hline 15 & howto & how-to & guide & tutorials
& help & how\_to \\ \hline 28 & games & game & timewaster & spiel &
jeu & bored \\ \hline 30 & java & python & perl & code & c++ &
delphi \\ \hline 39 & opensource & open\_source & open-source &
open.source & oss & foss \\ \hline 1152 & tobuy & wishlist & to\_buy
& buyme & wish-list & iwant \\ \hline
\end{tabular}
\vspace{1cm}
\centering \caption{Examples of most related tags measured by
Folkrank}
 \label{table-examples-folkrank}
\begin{tabular}{|c|c||c|c|c|c|c|} \hline
rank & tag & 1 & 2 & 3 & 4 & 5 \\ \hline\hline 13 & web2.0 & web &
ajax & tools & design & blog \\ \hline 15 & howto & reference &
linux & tutorial & programming & software \\ \hline 28 & games &
game & fun & flash & software & programming \\ \hline 30 & java &
programming & development & software & ajax & web \\ \hline 39 &
opensource & software & linux & programming & tools & web \\ \hline
 1152 & tobuy & toread & shopping & design & books & music \\ \hline
\end{tabular}
\end{table*}

Using each of the three measures introduced above,
we computed, for each of the $10,000$ most frequent tags of \del,
its most closely related tags.
Tables~\ref{table-examples-cooc}\,--\,\ref{table-examples-folkrank}
show a few selected examples. We observe that in many cases the cosine
similarity provides more synonyms than the other measures.
For instance, for tag \emph{web2.0} is returns some of its other
commonly used spellings.\footnote{The tag \emph{``web} at the fourth
position is likely to stem from some user who typed in \emph{``web
2.0''} which in the earlier del.icio.us was interpreted as two
separate tags \emph{``web} and \emph{2.0''}.}
For tag \emph{games}, the cosine similarity also provides tags that one
could consider as semantically \emph{similar} (like the singular
form \emph{game} or its German and French translations \emph{spiel}
and \emph{jeu}), while the other two measures provide \emph{related}
tags like \emph{fun} or \emph{software}. The same observation is
also made for the ``functional'' tag \emph{tobuy} (see
\cite{golder2006structure}), where the cosine similarity provides
tags with equivalent functional value, whereas the other measures
provide rather categories of items one could buy. An interesting
observation is also that \emph{java} and \emph{python} could be
considered as siblings in some suitable concept hierarchy.
A possible justification for these different behaviors is that the
cosine measure is measuring the frequency of co-occurrence with
other words \emph{in the global contexts}, whereas the co-occurrence
measure and --- to a lesser extent --- FolkRank
measure the frequency of co-occurrence with other words \emph{in the same
posts}. We will substantiate this assumption later in the paper on a
more general level.

\begin{table}
\caption{Overlap between the ten most closely related tags.}
 \label{table-overlap}\centering
\begin{tabular}{|c|c|c|} \hline
freq--folkrank & cosine--freq & cosine--folkrank \\ \hline\hline
6.7  & 1.7  & 1.1 \\
\hline\end{tabular} \end{table}
The first natural aspect to investigate is whether the most closely
related tags are shared across relatedness measures. We consider the
$10,000$ most popular tags in \del, and for each of them we compute
the $10$ most related tags according to each of the relatedness
measures. Table~\ref{table-overlap} reports the average number
of shared tags for the three relatedness measures.
We observe that relatedness by co-occurrence (freq) and by FolkRank
share a large fraction of the $10$ most closely related tags,
while the cosine relatedness displays little overlap with both of them.
 \begin{figure}
   \centering
   \includegraphics[width=0.6\columnwidth]{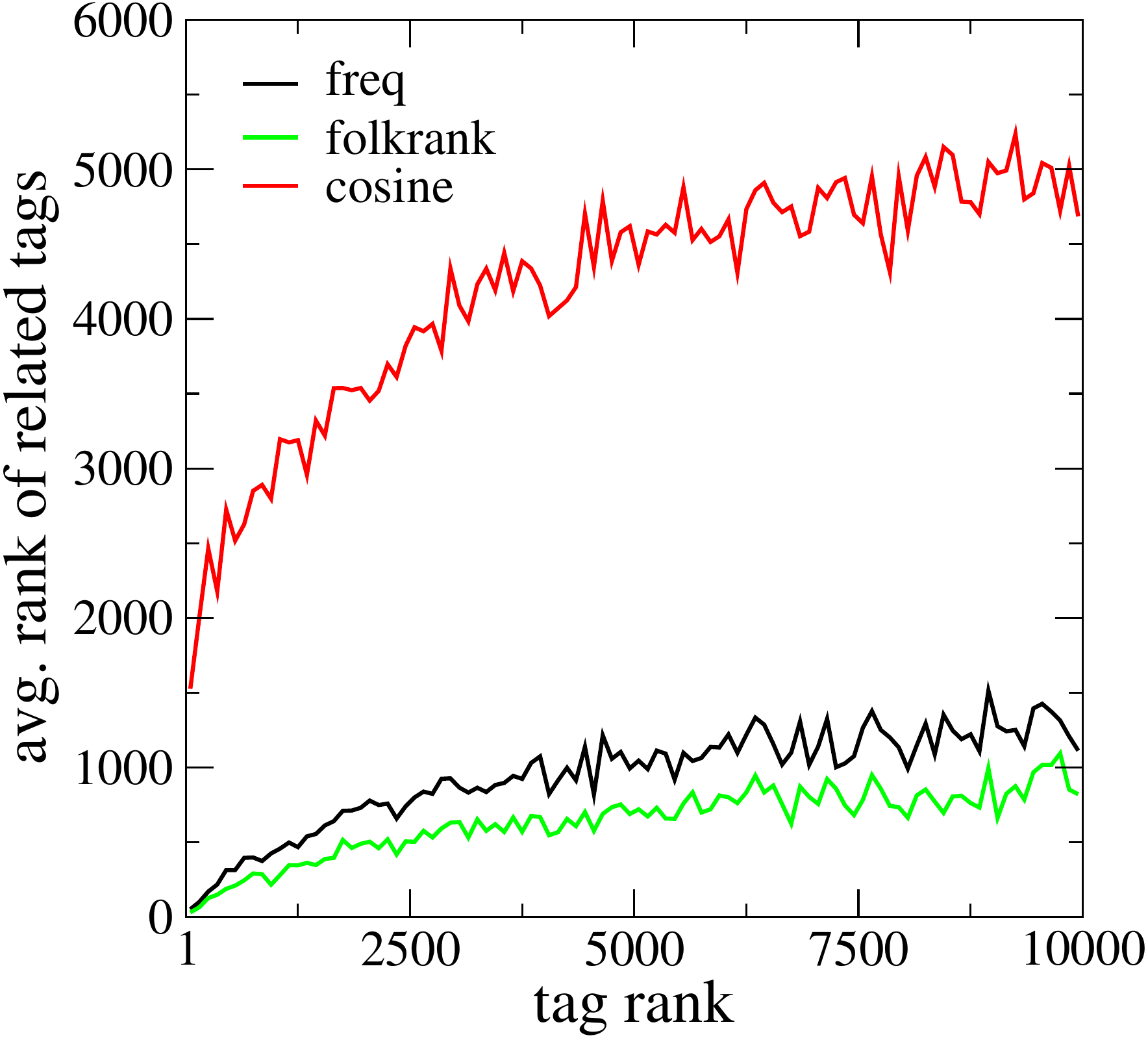}
   \caption{Average rank of the related tags
 as a function of the rank of the original tag.}\label{fig-rank}
 \end{figure}
To better investigate this point, we plot in Figure~\ref{fig-rank}
the average rank (according to global frequency) of the $10$
most closely related tags as a function of the rank of the original tag.
The average rank of the tags obtained by co-occurrence relatedness (black)
and by FolkRank (green) is low and increases slowly with the rank of the
original tag: this points out that most of the related tags are among
the high-frequency tags, independently of the original tag.
On the contrary, the cosine relatedness (red curve) displays a
different behavior: the rank of related tags increases
much faster with that of the original tag. That is, the tags
obtained from cosine-similarity relatedness belong to a broader class of tags,
not strongly correlated with rank (frequency).\footnote{Notice
that the curve for the cosine-similarity relatedness (red)
approaches a value of $~5\,000$ for high ranks: this is the value
one would expect if tag relatedness was independent from tag rank.
}

\section{Semantic Grounding}
 \label{sec-grounding}
In this section we shift perspective and move from the qualitative discussion 
of Section~\ref{sec-measures-II} to a more formal validation. Our strategy is 
to ground the relations between the original and the related tags by looking up 
the tags in a formal representation of word meanings.
As structured representations afford the definition of well-defined metrics of 
semantic similarity, one can investigate the type of \textit{semantic} 
relations that hold between the original tags and their related tags (obtained 
by using any of the relatedness measures we study).

In the following we ground our measures of tag relatedness by using WordNet~\cite{fellbaum98wordnet},
a semantic lexicon of the English language.
In WordNet words are grouped into \textit{synsets}, sets of synonyms
that represent one concept. Synsets are nodes in a network and links
between synsets represent semantic relations.

For nouns and verbs it is possible to restrict the
links in the network to (directed) \textit{is-a}
relationships only, so that a subsumption hierarchy can be defined.
The \textit{is-a} relation connects a \textit{hyponym} (more
specific synset) to a \textit{hypernym} (more general synset).
Since the \textit{is-a}
WordNet network for nouns and verbs consists of several disconnected
hierarchies, it is useful to add a (fake) global root node subsuming
all the roots of those hierarchies, making the graph fully connected
and allowing the definition of several graph-based similarity
metrics between pairs of nouns and pairs of verbs. We will use such
measures to ground our tag-based measures of relatedness in
folksonomies.

We measure the similarity in WordNet
using both the taxonomic shortest-path length and a distance measure
introduced by Jiang and Conrath~\cite{jiang97semantic} that combines
the taxonomic path length with an information-theoretic similarity measure by Resnik~\cite{resnik95using}.
We use the implementation of those measures available
in the WordNet::Similarity
library~\cite{university04wordnetsimilarity}.
We remark that \cite{budanitsky06evaluating} provides a pragmatic grounding of the
Jiang-Conrath measure by means of user studies and by its
superior performance in the correction of spelling errors.
This way, our semantic grounding in WordNet of the
folksonomy similarity measures is extended to a pragmatic grounding
in the experiments of \cite{budanitsky06evaluating}.

The program outlined above is only viable if a significant fraction of the 
popular tags in \del is also present present in WordNet.
Several factors limit the WordNet coverage of \del tags: WordNet only covers 
the English language and contains a static body of words, while \del contains 
tags from different languages and is an open-ended system. This is not a big 
problem in practice because, to date, the vast majority of \del tags are 
grounded in the English language. Another limiting factor is the structure of 
WordNet itself, where the measures described above can only be implemented for 
nouns and verbs, separately. Many tags are actually 
adjectives~\cite{golder2006structure} and although their grounding 
is possible no distance based on the subsumption hierarchy can be computed 
in the adjective partition of WordNet. Nevertheless, the nominal form of the 
adjective is often covered by the noun partition.
Despite this, if we consider the popular tags in \del, a significant fraction 
of them is actually covered by WordNet: Roughly 61\% of the $10\,000$ most 
frequent tags in \del can be found in WordNet. In the following, to make 
contact with the previous sections, we will focus on these tags.

\begin{table}
 \caption{Average semantic distance, measured in
WordNet, from the original tag to the most closely related one.
\label{table-semantic-distance} }
\centering
\begin{tabular}{|r|c|c|c|} \hline
similarity metric & freq & folkrank & cosine \\ \hline\hline
shortest path & 7.4 & 7.8 & 6.3 \\ \hline
Jiang-Conrath & 13.1  & 13.6  & 10.8 \\
\hline\end{tabular}\end{table}
A first assessment of the measures of relatedness can be carried out by
measuring -- in WordNet -- the average semantic distance between a
tag and the corresponding most closely related tag
according to each one of the relatedness measures we consider.
Given a measure of relatedness, we loop over the tags that
are both in \del and WordNet, and for each of those tags
we use the chosen measure of relatedness to find the corresponding
most related tag. If the most related tag is also in WordNet,
we measure the semantic distance between the synsets
that contain the original tag and the most closely related tag, respectively.
In the case of the shortest-path distance, if any of the tags occurs in
more than one synset, we select synsets which minimizes the path length.
Table~\ref{table-semantic-distance} reports the average semantic
distance, computed in WordNet by using both the (edge) shortest-path
length and the Jiang-Conrath distance. The cosine relatedness points
to tags that are semantically closer according to both measures. We
remark once more that the Jiang-Conrath measure has been validated
in user studies~\cite{budanitsky06evaluating}, so that
Table~\ref{table-semantic-distance} actually deals with distances
cognitively perceived by human subjects.
The closer semantic proximity of tags obtained by cosine relatedness
was intuitively apparent from the comparison of Table~\ref{table-examples-cosine}
with Table~\ref{table-examples-cooc} and Table~\ref{table-examples-folkrank},
but now we are able to ground this statement through user-validated measures
based on the subsumption hierarchy of WordNet.

As noted in Section~\ref{sec-measures-II}, the tags obtained
via the cosine-similarity relatedness measure appear to be ``synonyms'' or
``siblings'' of the original tag, while the two other measures of
relatedness seem to provide ``more general'' tags. The possibility
of locating tags in the WordNet hierarchy allows us to be more
precise about the nature of these relations. In the rest of this
section we will focus on the shortest paths in WordNet that lead
from an initial tag to its most closely related tag (according to
the different similarity measures), and characterize the length and
edge composition (hypernym/hyponym) of such paths.

\begin{table}
 \caption{Probabilities of the lengths of the shortest path leading 
 from the original tag to the most closely related one. Path lengths are 
 computed using the subsumption hierarchy in WordNet.
\label{table-pathlen}}
 \centering \begin{tabular}{|r|c|c|c|c|} \hline shortest path length & 0 & 1 & 
 2 & $\geq 3$ \\ 
 \hline\hline freq & 0.05 & 0.04 & 0.06 & 0.85\\ \hline 
 folkrank & 0.04 & 0.04 & 0.05 & 0.87\\ \hline
 cosine & 0.18 & 0.03 & 0.09 & 0.70\\
 \hline\end{tabular}\end{table}

Table~\ref{table-pathlen} summarizes the probabilities of the shor\-test-path 
lengths $n$ (number of edges) connecting a tag to its closest 
related tag in WordNet. The FolkRank and co-occurrence relatedness have similar 
probabilities. The cosine relatedness displays higher values
at $n=0$ and $n=2$ and a comparatively depleted number of paths with $n=1$.
The higher value at $n=0$ is due to the detection 
of actual synonyms; \ie the cosine relatedness, in about $18$\,\% of the cases, 
points to a tag which belongs to the same synset of the original tag. The 
smaller number of paths with $n=1$ (one single edge in WordNet) is 
consistent with the idea that the cosine relatedness favors siblings/synonymous 
tags: moving by a single edge, instead, leads to either a hypernym or a hyponym 
in the WordNet hierarchy, never to a sibling. The higher value at $n=2$ (paths with two 
edges in WordNet) may be compatible with the sibling relation, but in order to 
ascertain it we have to characterize the average edge composition of these 
paths.

\begin{figure}
  \center
  \includegraphics[width=0.9\columnwidth]{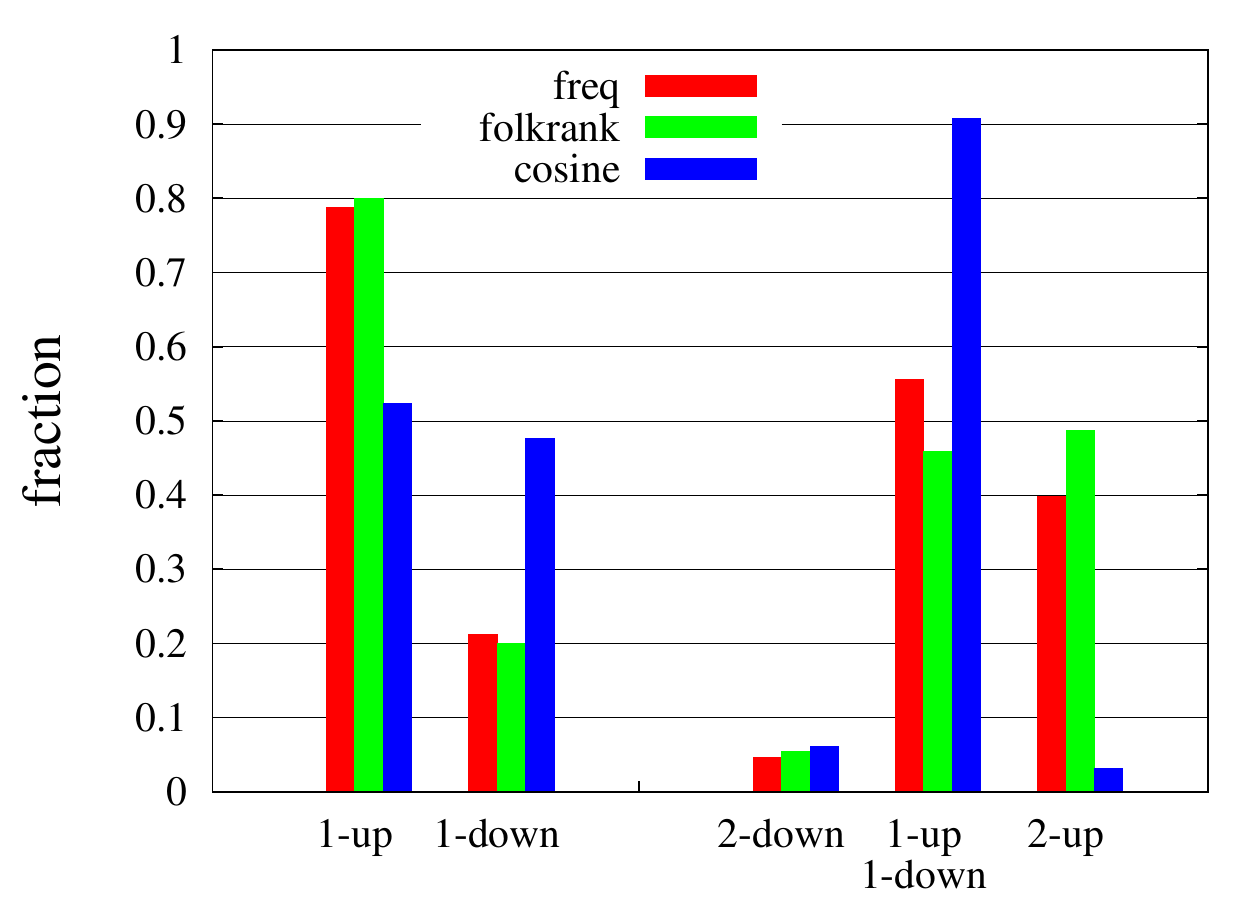}
  \caption{Edge composition of the shortest paths of length $1$ (left)
and $2$ (right).  An ``up'' edge leads to a hypernym,
while a ``down'' edge leads to a hyponym.
}
\label{fig-path-composition2}
\end{figure}
Figure~\ref{fig-path-composition2} displays the average edge type
composition (hypernym/hyponym edges) for paths of length $1$ and
$2$.
For the cosine-similarity relatedness (blue), we observe that the paths with $n=2$ (right-hand side of Figure~\ref{fig-path-composition2}) consist almost entirely ($90$\%) of one hypernym edge (up) and one hyponym edge (down), \ie these paths do lead to siblings. Notice how the path composition is very
different for the other relatedness measures: in those cases roughly
half of the paths consist of two hypernym edges in the WordNet
hierarchy. We observe a similar behavior for $n=1$, where the cosine
relatedness has no statistically preferred direction, while the
other measures of relatedness point preferentially to hypernyms.

\section{Discussion and Perspectives}
 \label{sec-discussion}

The main contribution of this paper is a methodological one. Several measures 
of relatedness have been proposed in the literature, but given the fluid and 
open-ended nature of social bookmarking systems, it is hard to characterize -- 
from the semantic point of view -- what kind of relations they establish. As 
these relations constitute an important building block for extracting 
formalized knowledge, a deeper understanding of these
 measures is needed. Here we proposed to ground different measures of tag 
relatedness in a folksonomy by mapping \del tags, when possible, on WordNet 
synsets and using well-established measures of semantic distance in WordNet to 
gain insight into their respective characteristics.

Our results can be taken as indicators that the choice of an appropriate 
relatedness measure is able to yield valuable input for learning semantic term 
relationships from folksonomies. We will close by briefly discussing which of 
the three relatedness measures we studied is best for \dots

\begin{itemize}

\item \emph{\dots synonym discovery.} The cosine similarity is clearly the 
measure to choose when one would like to discover synonyms. As shown in this 
work, cosine similarity delivers not only spelling variants but also terms that 
belong to the same WordNet synset. 

\item \emph{\dots concept hierarchy.} Both FolkRank and co-occurrence 
relatedness seemed to yield more general tags in our analyses. This is why we 
think that these measures provide valuable input for algorithms to extract 
taxonomic relationships between tags. 

\item \emph{\dots discovery of multi-word lexemes.} Depending on the allowed 
tag delimiters, it can happen that multi-word lexemes end up as several tags. 
Our experiment indicates that FolkRank is best to discover these cases. For the 
tag \emph{open}, for instance, it is the only of the three algorithms which has 
\emph{source} within the ten most related tags and vice 
versa.\footnote{\emph{Open} is at position 6 for \emph{source}, and 
\emph{source} is at position 3 for \emph{open}.} 

\end{itemize}

Future work includes the analysis of further relatedness measures, \eg based on 
representations in the vector spaces spanned by the users or resources. We are 
furthermore currently working on adapting existing ontology learning techniques 
to folksonomies, including the presented measures.

\subsection*{Acknowledgment}
This research has been partly supported by the TAGora project
(\texttt{FP6-IST5-34721}) funded by the Future and Emerging
Technologies program (IST-FET) of the European Commission. We thank
A.~Baldassarri, V.~Loreto, F.~Menczer, V.~D.~P.~Servedio and L.
Steels for many stimulating discussions.


\begin{thebibliography}{10}

\bibitem{brin98anatomy}
Sergey Brin and Lawrence Page, `{T}he {A}natomy of a {L}arge-{S}cale
  {H}ypertextual {W}eb {S}earch {E}ngine', {\em Computer Networks and ISDN
  Systems}, {\bf 30}(1-7),  107--117, (April 1998).

\bibitem{budanitsky06evaluating}
Alexander Budanitsky and Graeme Hirst, `Evaluating wordnet-based measures of
  lexical semantic relatedness', {\em Computational Linguistics}, {\bf 32}(1),
  13--47, (2006).

\bibitem{cattuto2007pnas}
Ciro Cattuto, Vittorio Loreto, and Luciano Pietronero, `Semiotic dynamics and
  collaborative tagging', {\em Proceedings of the National Academy of Sciences
  (PNAS)}, {\bf 104},  1461--1464, (2007).

\bibitem{cattuto2007aicomm}
Ciro Cattuto, Christoph Schmitz, Andrea Baldassarri, Vito D.~P. Servedio,
  Vittorio Loreto, Andreas Hotho, Miranda Grahl, and Gerd Stumme, `Network
  properties of folksonomies', {\em AI Communications Journal, Special Issue on
  Network Analysis in Natural Sciences and Engineering}, {\bf 20}(4),
  245--262, (2007).

\bibitem{ciminao06phd}
Philipp Cimiano, {\em Ontology Learning and Population from Text ---
  Algorithms, Evaluation and Applications}, Springer, Berlin--Heidelberg,
  Germany, 2006.
\newblock Originally published as PhD Thesis, 2006, Universität Karlsruhe (TH),
  Karlsruhe, Germany.

\bibitem{fellbaum98wordnet}
{\em {WordNet: an electronic lexical database}}, ed., Christiane Fellbaum, MIT
  Press, 1998.

\bibitem{firth57synopsis}
J.~R. Firth, `A synopsis of linguistic theory 1930-55.', {\em Studies in
  Linguistic Analysis (special volume of the Philological Society)}, {\bf
  1952-59},  1--32, (1957).

\bibitem{Ganter99}
B.~Ganter and R.~Wille, {\em Formal Concept Analysis: Mathematical
  Foundations}, Spring-Verlag, 1999.

\bibitem{golder2006structure}
Scott Golder and Bernardo~A. Huberman, `The structure of collaborative tagging
  systems', {\em Journal of Information Science}, {\bf 32}(2),  198--208,
  (April 2006).

\bibitem{Halpin_et_al_2006}
H.~Halpin, V.~Robu, and H.~Shepard, `The dynamics and semantics of
  collaborative tagging', in {\em Proceedings of the 1st Semantic Authoring and
  Annotation Workshop (SAAW'06)}, (2006).

\bibitem{harris68mathematical}
Z.~S. Harris, {\em Mathematical Structures of Language}, Wiley, New York, 1968.

\bibitem{heymann06collaborative}
Paul Heymann and Hector Garcia-Molina, `Collaborative creation of communal
  hierarchical taxonomies in social tagging systems', Technical Report 2006-10,
  Computer Science Department, (April 2006).

\bibitem{hotho06information}
Andreas Hotho, Robert J\"{a}schke, Christoph Schmitz, and Gerd Stumme,
  `Information retrieval in folksonomies: Search and ranking', in {\em The
  Semantic Web: Research and Applications}, eds., York Sure and John Domingue,
  volume 4011 of {\em LNAI}, pp. 411--426, Heidelberg, (2006). Springer.

\bibitem{jiang97semantic}
Jay~J. Jiang and David~W. Conrath, `{Semantic Similarity based on Corpus
  Statistics and Lexical Taxonomy}', in {\em Proceedings of the International
  Conference on Research in Computational Linguistics (ROCLING)}. Taiwan,
  (1997).

\bibitem{lambiotte05tripartite}
R.~Lambiotte and M.~Ausloos, `Collaborative tagging as a tripartite network',
  {\em Lecture Notes in Computer Science}, {\bf 3993},  1114, (Dec 2005).

\bibitem{LehmannWille95}
F.~Lehmann and R.~Wille, `A triadic approach to formal concept analysis', in
  {\em Conceptual Structures: Applications, Implementation and Theory}, eds.,
  G.~Ellis, R.~Levinson, W.~Rich, and J.~F. Sowa, volume 954 of {\em Lecture
  Notes in Computer Science}. Springer, (1995).

\bibitem{mathes04folksonomies}
Adam Mathes.
\newblock {F}olksonomies -- {C}ooperative {C}lassification and {C}ommunication
  {T}hrough {S}hared {M}etadata, December 2004.
\newblock
  http://www.adammathes.com/academic/computer-mediated-communication/folksonom%
ies.html.

\bibitem{mika2005ontologies}
Peter Mika, `Ontologies are us: A unified model of social networks and
  semantics', in {\em International Semantic Web Conference}, LNCS, pp.
  522--536. Springer, (2005).

\bibitem{mohammadSubmittedDistributional}
Saif Mohammad and Graeme Hirst.
\newblock Distributional measures as proxies for semantic relatedness.
\newblock Submitted for publication,
  \url{http://ftp.cs.toronto.edu/pub/gh/Mohammad+Hirst-2005.pdf}.

\bibitem{page98pagerank}
L.~Page, S.~Brin, R.~Motwani, and T.~Winograd, `The {P}age{R}ank citation
  ranking: Bringing order to the web', in {\em WWW'98}, pp. 161--172, Brisbane,
  Australia, (1998).

\bibitem{university04wordnetsimilarity}
T.~Pedersen, S.~Patwardhan, and J.~Michelizzi.
\newblock Wordnet::similarity - measuring the relatedness of concepts, 2004.
\newblock http://citeseer.ist.psu.edu/665035.html.

\bibitem{resnik95using}
Philip Resnik, `{Using Information Content to Evaluate Semantic Similarity in a
  Taxonomy}', in {\em Proceedings of the XI International Joint Conferences on
  Artificial}, pp. 448--453, (1995).

\bibitem{salton89}
Gerard Salton, {\em Automatic text processing: the transformation, analysis,
  and retrieval of information by computer}, Addison-Wesley Longman Publishing
  Co., Inc., Boston, MA, USA, 1989.

\bibitem{schmitz2006inducing}
Patrick Schmitz, `Inducing ontology from {F}lickr tags.', in {\em Collaborative
  Web Tagging Workshop at WWW2006, Edinburgh, Scotland}, (May 2006).

\bibitem{zhang06-emergent}
Lei Zhang, Xian Wu, and Yong Yu, `Emergent semantics from folksonomies: A
  quantitative study', {\em Journal on Data Semantics VI}, (2006).

\end{thebibliography}
\end{document}